# Double charge exchange ($^{11}$B,$^{11}$Li) reaction for double beta decay response

K. Takahisa[a], H. Ejiri[a], H. Akimune[b], H.Fujita[a], R.Matumiya[a], T.Ohta[a], T.Shima[a], M. Tanaka[c] and M. Yosoi[a]

[a]Research Center for Nuclear Physics, Osaka University, Osaka, 567-0047, Japan,
[b]Konan University, Hyogo 658-8501, Japan
[c]Kobe Tokiwa University, Kobe 653-0838, Japan

Abstract

The ($^{11}$B,$^{11}$Li) double charge-exchange reaction (DCER) at $E(^{11}B)/A$=80 MeV was measured for the first time to demonstrate the feasibility of the reaction for studying neutrino nuclear responses for double beta decays (DBD). The $^{13}$C($^{11}$B,$^{11}$Li)$^{13}$O reaction shows strengths at the ground state and low and high excitation giant resonance regions. The $^{56}$Fe ($^{11}$B,$^{11}$Li) $^{56}$Ni reaction shows the large strengths in the possible double giant resonance region and beyond, but shows no strengths in the low excitation region below 5 MeV, suggesting strong concentration of the DBD strength at the high excitation region. The DCER is used to evaluate the spin isospin strengths relevant to DBD responses.

Keywords: Double charge-exchange reaction, Double beta decays

## 1., Introduction

Double beta decays (DBD) are of current interest in view of particle, astro and nuclear physics[1,2]. Neutrino-less double beta decays (0νββ) with lepton number violation of $\Delta L$=2, are sensitive to the Majorana masses of light and heavy neutrinos(ν), right-left mixings of weak currents, SUSY-neutrino couplings, and so on beyond the standard model. These are discussed in recent review articles and referenced therein[1,2]. The nuclear response (square of the DBD nuclear matrix element NME) are crucial to extract the Majorana neutrino mass and the neutrino properties from the experimental DBD rate, if it is observed.

Here the responses are mainly of spin isospin (στ) responses. The στ strengths are pushed up by the strong repulsive στ interactions to the highly excited GT (Gamow Teller) giant resonance (GTGR), leaving little strengths in the low-lying states, and likewise double τσ strengths may be considered to be located in the highly excited region, leaving little strengths at the low-excitation region[3,4,5]. Actually, the GT NMEs are reduced by a factor $g_A^{eff}/g_A$~0.25 with respect to the single quasi-particle NMEs. In fact the double GT NMEs derived from the 2 neutrino DBD rates are shown to be reduced by the factor $(g_A^{eff}/g_A)^2$~0.05.

So far, the single charge-exchange ($^3$He,t) reactions at $E(^3He)/A$=150MeV have been extensively used to study the single τσ responses because the medium energy projectile excite preferentially the τσ mode[1,2].

The double charge exchange reactions (DCERs) are interesting to study the DBD responses. Actually, the DCERs of (π$^+$,π$^-$) have been measured to study the GDR*IAS (Giant Dipole Resonance built on isobaric analog states) and DIAS (Double Isobaric Analogue State) and DGDR (Double Giant Dipole Resonance)[6]. Since pions are spin-zero particles, spin-flip transitions are strongly suppressed. Heavy ion (HI) DCERs of $^{24}$Mg($^{18}$O,$^{18}$Ne)$^{24}$Ne reaction[7] were measured at 100 MeV/$A$ at NSCL-MSU and 76 MeV/$A$ at GANIL[8]. However, there were large statistical uncertainties. Recently, the DCER of $^{40}$Ca($^{18}$O,$^{18}$Ne)$^{40}$Ar was measured at $E/A$~15MeV[9,10]. Here, the isospin of target nucleus is

$T_z=0$ with $N=Z$ and the projectile energy is well below 100 MeV, and thus no strong $\tau\sigma$ excitations at the high excitation region are expected. Note that the DBD nuclei of current interest are $N \gg Z$.

It is important to use medium-energy ($E/A \sim 100$ MeV) light HIs for direct excitation of the $\tau\sigma$ mode. The present work aims at demonstrating experimental feasibility of the lightest HI DCER of ($^{11}$B,$^{11}$Li) at $E(^{11}$B$)/A=80$ MeV for studying $\tau\sigma$ strength distributions relevant to the DBD response in medium-mass nuclei.

## 2. Experimental Method

In the present experiment, we studied, for the first time, the DCER of ($^{11}$B,$^{11}$Li) with the lightest ($A=11$) heavy ion projectile and the medium energy ($E/A=80$ MeV) to minimize the distortion effect at this energy region and to ensure the dominance of the $\tau\sigma$ interaction. This is the lightest stable heavy-ion DCER with charged out-going particles.

Research and development of the ($^{11}$B,$^{11}$Li) reaction was made first at RCNP by using of NEOMAFIOS[11,12] ECR(Electron Cyclotron Resonance) ion source. Recently a new ion source has been developed to produce higher intensity beam[13,14]. The 18 GHz superconducting ECR ion source has been installed in order to increase accelerated beam intensities for various heavy ions. In the present experiment, stable boron beams($^{11}$B$^{5+}$) have also been produced by the MIVOC method[14]. The targets used are the light and medium nuclei of $^{13}$C and $^{56}$Fe.

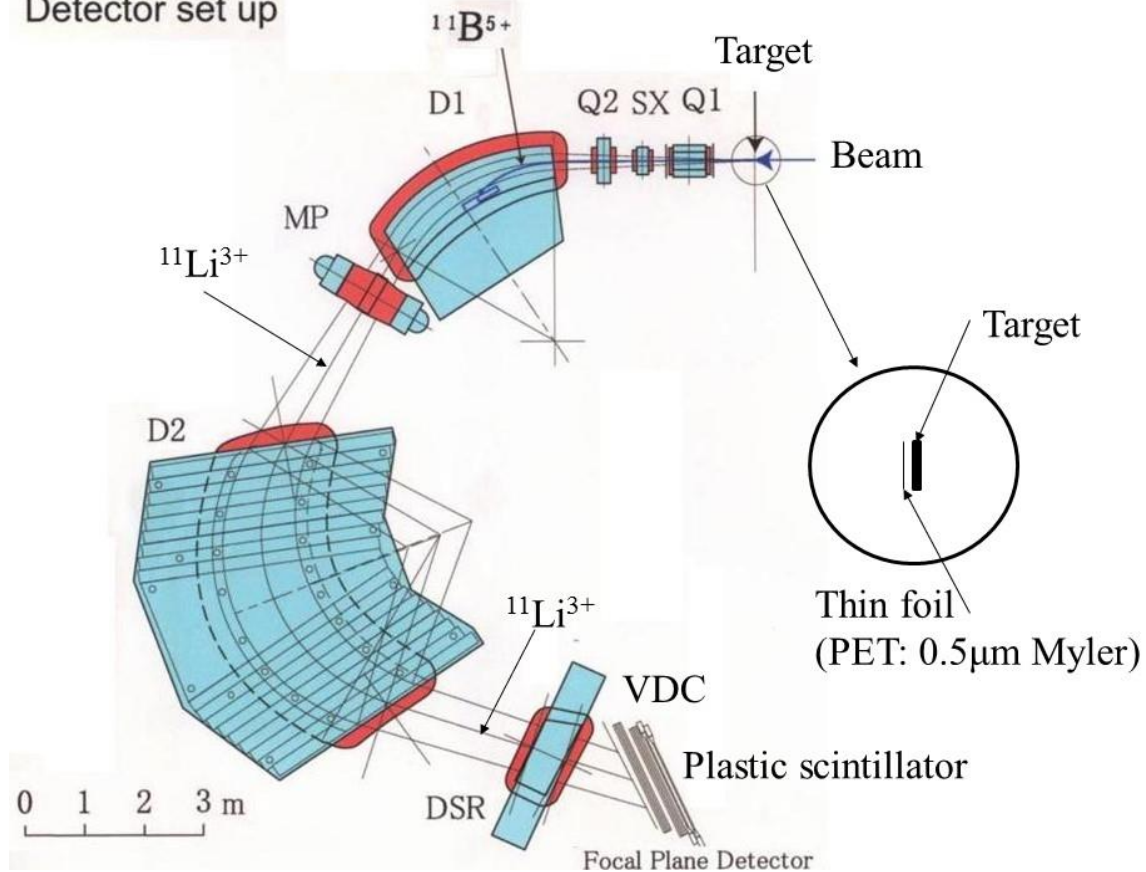

Figure 1. The Grand Raiden spectrometer. Q1, Q2: quadrupole  D1:  DSR: for DCER($^{11}$B,$^{11}$Li)

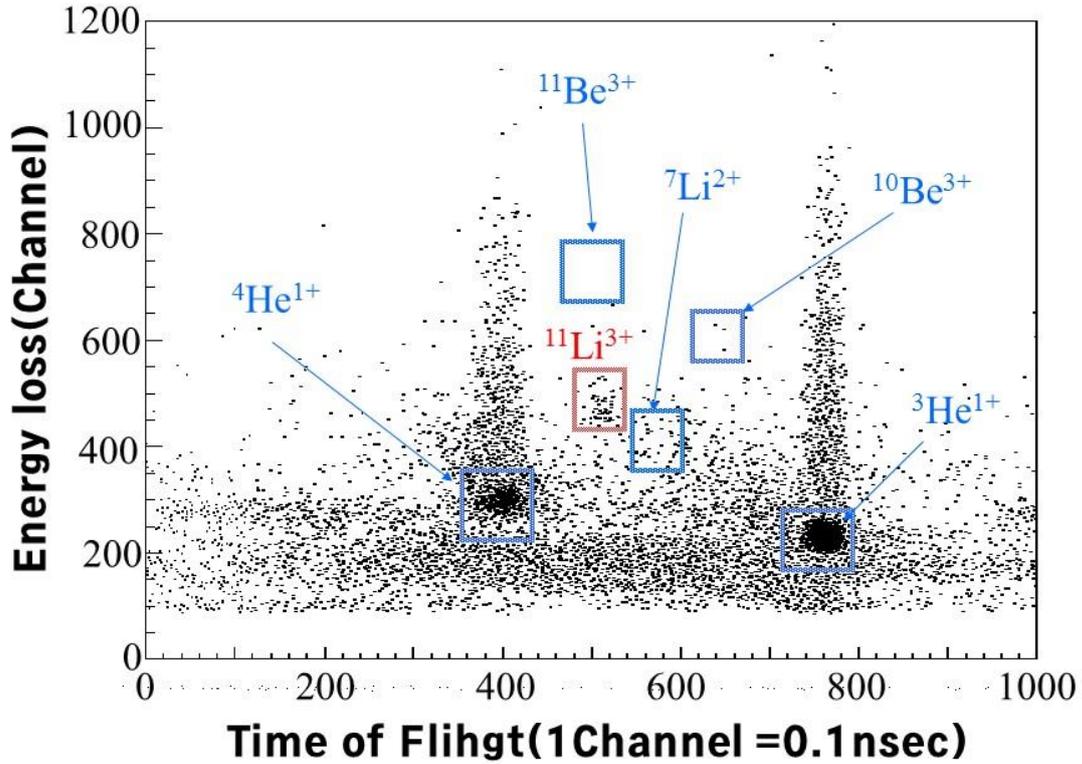

Figure 2. Particle identification of the scattered $^{11}$Li particle by using TOF and the energy loss at plastic scintillator (thickness:3mm). The start signal of the TOF is RF signal from the Ring cyclotron.

The spectrometer Grand Raiden[15] was set at 0 degree. The horizontal and vertical opening angles were 30 m radian and the full solid angle were 1.6 m steradian. It covers the angular range of 0 – 1.7 degree as shown in Fig.1. The $^{11}$B$^{5+}$ beam is collected by the internal Faraday cup in the first dipole (D1) magnet. A thin foil (PET: 0.5 $\mu$m Mylar) was placed just after the target foil to reduce the $^{11}$B$^{3+}$ ions as shown in Fig.1. We could clearly identify the scattered $^{11}$Li particle by using TOF (Time Of Flight) and the energy loss at the plastic scintillator with 3mm in thickness at the focal plane, as shown in Fig.2. The start signal for the TOF is the RF signal of the Ring cyclotron. The experimental background was very low.

The momentum transfer for this ($^{11}$B,$^{11}$Li) CER at the present angle of θ = 0° - 2° is around 160MeV/c for the ground state, and 180 MeV/c for the excitation region at $E$=20-30MeV. The corresponding angular momentum transfer is $\Delta l$=1～3ℏ.

## 3. Cross-sections for the $^{13}$C($^{11}$B,$^{11}$Li)$^{13}$O reaction and the $^{56}$Fe($^{11}$B,$^{11}$Li) $^{56}$Nireaction.

To check the feasibility of the DCE the $^{13}$C ($^{11}$B,$^{11}$Li)$^{13}$O reaction was measured by using a $^{13}$C target with 2.8 mg/cm$^2$. The sharp peak for the $^{13}$O ground state transition was clearly observed, as shown in Fig.3. The cross section is 50.9±7.6 nb/sr. The energy calibration of the spectrometer has been made by referring to this peak. The energy resolution of 1.2 MeV is partly due to the beam resolution (0.5MeV) and the difference of energy losses in the target (0.76MeV). In the high excitation region above the cross section increases as the excitation energy increases. The cross section at the low excitation region of E=0-5 MeV is around 20 nb/sr/MeV, while the cross section at the high excitation region of E=20-30 MeV is around 100 nb /sr/MeV. Broad bumps around E=30 MeV and 45 MeV may correspond to the giant resonance built on isobaric analog states GDR*IAS and the double giant resonance DGDR observed in the double charge exchange $(\pi^+,\pi^-)$ reaction[4] on $^{13}C$ .

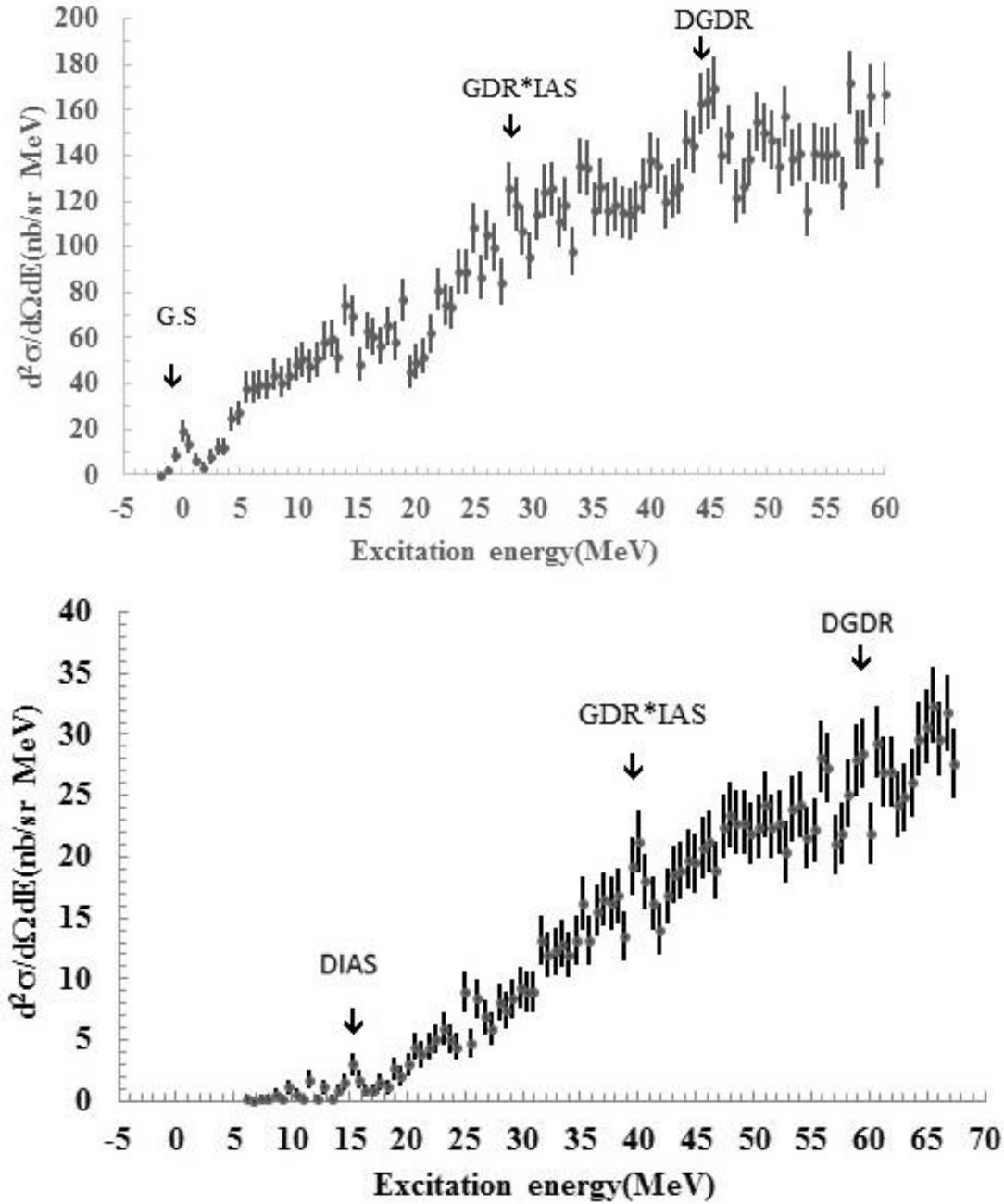

Figure 3. Cross-sections for the $^{13}C(^{11}B,^{11}Li)^{13}O$ reaction(upper) and the $^{56}Fe(^{11}B,^{11}Li)$ $^{56}Ni$ reaction(low).

Then we measured the $(^{11}B,^{11}Li)$ reaction on $^{56}Fe$ with N-Z=4 ($T_Z$=2) (99% enriched in $^{56}Fe$, 11mg/cm$^2$). So far the $^{56}Fe(\pi^+,\pi^-)$ ($\Delta T$=2,$\Delta S$=0) reaction has been measured[3]. In this reaction, DIAS and GDR*IAS were observed. In the present case of the $^{56}Fe(^{11}B,^{11}Li)^{56}Ni$ ($\Delta T$=2) reaction, it is possible to excite not only DIAS and GDR*IAS, but also DGTR. A theoretical calculation of the DGTR for $^{56}$Fe is available. The DGTR strength was estimated to be concentrated around 25.6MeV.

The energy spectrum of the $^{56}$Fe($^{11}$B,$^{11}$Li)$^{56}$Ni reaction is shown in Fig.3. No clear peaks at the low excitation region of E=0-5 MeV in $^{56}$Ni were observed. The cross section in this low excitation region is less than 0.06 nb/sr/MeV. The double isobaric analogue state (DIAS) with the cross section of 111±2nb/sr was observed broad bumps at the excitation regions of 40 MeV and 60 MeV in Fig.3(bottom) may correspond to GDR*IAS and DGDR respectively, as observed in the ($\pi^+,\pi^-$) [5], while the bump at around E=25 MeV might include the DGTR strength although one need further experimental study to confirm them.

The DCER cross section σ(H) in the high excitation region of E=20-30 MeV is 7 nb/sr/MeV, while the cross section σ(L) in the low excitation region of E=0-5MeV is less than 0.06 nb/sr/MeV. The ratio of σ(L)/ σ(H) is < 0.01. In case of the single CER, the ratio is around 0.1. Thus the DCE strengths are pushed up in the high GR region, and the strengths to the low lying states are deduced at least by two orders of magnitudes. This may suggest a severe deduction of the DBD strengths as observed in case of the 2νββ strengths[3].

### 4.Conclusion

The present work demonstrates for the first time experimental feasibility of the ($^{11}$B,$^{11}$Li) reaction at 80MeV/nucleon. In the case of the $^{13}$C($^{11}$B,$^{11}$Li)$^{13}$O reaction, the ground and excited states as well as GDR*IAS and DGDR were observes. This shows good correspondence with the ($\pi^+,\pi^-$) data on GDR*IAS and DGDR. The broad peak around 25MeV may be the DGTR as suggested theoretically. The ($^{11}$B,$^{11}$Li) reaction are good probes for DGDR.

The $^{56}$Fe($^{11}$B,$^{11}$Li)$^{56}$Ni reaction shows the DIAS and GDR*IAS and strength at DGTR region. However, it does not show any strength at the low excitation region.

The τσ interaction of $V_{\tau\sigma}$ is dominant at the present beam energy of 80MeV/nucleon. The present data on $^{56}$Fe show the strong concentration of the τσ strengths at the high excitation region and severe reduction of the τσ strength at low excitation. The present DCER with the lightest medium-energy projectile with 80MeV/nucleon is used to study spin isospin strengths relevant to double beta decay,.

### Acknowledgements

This experiments was performed at RCNP under the Experimental Programs E241. The authors thank the RCNP accelerator group for their effort in providing a high-quality $^{11}$B beam, and Dr.Yorita for development of the boron ion source.